\documentclass[twocolumn,prl,superscriptaddress,noshowpacs,epsf]{revtex4}

\usepackage{graphicx}

\begin{document}
\title{Superconducting Circuits and Quantum Information\footnote{J.Q. You 
(jqyou@fudan.edu.cn) is a professor of physics at
Fudan University in Shanghai, China.
Franco Nori (nori@umich.edu) is a professor of physics and applied physics
at the University of Michigan in Ann Arbor
and the director of the Digital Materials Laboratory 
at Japan's Institute of Physical and Chemical Research (RIKEN).}
}

%
\author{J. Q. You}
%
\author{Franco Nori}

\begin{abstract}

Superconducting circuits can behave like atoms making transitions between two levels.
Such circuits can test quantum mechanics at macroscopic scales and be used to conduct
atomic-physics experiments on a silicon chip.
\end{abstract}

\pacs{03.67.Pp, 74.50.+r}
\maketitle

Quantum bits, or qubits, form the heart of quantum-information processing schemes.
Because of the quantum parallelism and entanglement that arise from the superposition 
of states in two-level qubit systems, researchers expect eventual quantum computers 
to tackle tasks, such as factoring large numbers and simulating large 
quantum systems, that no ordinary computers can do in a practical time frame.

Quantum computing involves preparing, manipulating, and reading out the quantum 
states of a many-qubit system. So it is desirable to have qubits 
that can be individually controlled. Moreover, they should be scalable; 
that is, simply adding more qubits should create a larger circuit capable of 
more complex calculations. Solid-state qubits satisfy these requirements.

Fortunately, very small solid-state devices can behave quantum mechanically.
As the size of a bulk conductor becomes increasingly smaller, its quasi-continuous
electron conduction band turns into discrete energy levels.
An example is a quantum dot, in which electrons are confined to a small semiconducting 
or metallic box or island composed of millions of atoms. The problem is that
the electron states of that island quickly decohere
as the microscopic degrees of freedom strongly interact with the environment. 
A bulk superconductor, in contrast, is composed of many paired electrons that
condense into a single-level state. 
This superconducting state involves macroscopic degrees of freedom and thus
exhibits better quantum coherence. 
By reducing the size of the superconductor, one can reduce the coupling of the 
superconducting state to the environment and thereby further improve 
the quantum coherence. 

Various experiments on superconducting circuits have demonstrated 
as much~\cite{NEC,Saclay,Stony,Delft,KB} and those schemes are regarded
as promising candidates of qubits that can process quantum information
(see Physics Today, June 2002, page 14).
Not surprisingly, there is a deep analogy between natural atoms and the artificial 
atoms composed of electrons confined in small superconducting islands.
Both have discrete energy levels and exhibit coherent quantum oscillations  
between those levels{\bf ---}so-called Rabi oscillations. But whereas natural atoms 
are driven using visible or microwave photons that excite electrons from one state 
to another, the artificial atoms in the circuits are driven by currents, voltages, 
and microwave photons. The resulting electric and magnetic fields control the 
tunneling of electrons between the superconducting island and nearby electrodes. 
The effects of those fields on the circuits are 
the analogues of the Stark and Zeeman effects in atoms.

\vspace{.25cm}\noindent \fbox{\parbox{8.4cm}{

{\bf\large Box~1. Parameters of Superconducting Qubits}


\begin{center}
\begin{tabular}{l|c|c|c|c}
\hline\hline
& Charge
& Charge-flux
& Flux
& Phase
\\
\hline
$E_J/E_c$ & $0.1$ & 1 & $10$ & $10^6$\\
$\nu_{01}$ & $10$~GHz & $20$~GHz & $10$~GHz & $10$~GHz\\
$T_1$ & $1-10$~$\mu$s & $1-10$~$\mu$s & $1-10$~$\mu$s & $1-10$~$\mu$s\\
$T_2$ & $0.1-1$~$\mu$s & $0.1-1$~$\mu$s & $1-10$~$\mu$s & $0.1-1$~$\mu$s\\
\hline\hline
\end{tabular}
\end{center}


~~~The ratio of two energy sclaes{\bf ---}the Josephson coupling energy $E_J$ 
and the charging energy $E_c${\bf ---} determines whether the phase or the charge 
dominates the behavior of the qubit. Moreover, a low enough temperature $T$ 
($k_BT$ smaller than the level splitting of the qubit) 
prevents the qubit states from thermally smearing.

~~~The values listed in the table are approximate orders of magnitude from recent 
experiments with different cicuits.
Here, $h\nu_{01}$ is the level splitting of the qubit (that is, the energy-level 
difference of the two lowest states $E_1-E_0$), which depends on the applied bias. 
$T_{1}$ is the average time that the system takes for its excited state 
$|1\rangle$ to decay to the ground state $|0\rangle$.
$T_{2}$ represents the average time over which the qubit energy-level difference
does not vary.
The relaxation and decoherence times, $T_1$ and $T_2$, are strongly affected by
the environment of the artificial atom.
The readout visibility $V$, defined as the maximum qubit population 
difference observed in a Rabi oscillation or Ramsey fringe,
can reach more than $96\%$
~\cite{NEC}.
The coherence quality factor $Q=\pi T_2\nu_{01}$ is roughly $10^5$, 
the number of one-qubit operations achievable before the system 
decoheres~\cite{Saclay}. 
}}
\vspace{.1cm}\noindent

Differences between quantum circuits and natural atoms include how strongly 
each system couples to its environment; the coupling is weak for atoms and 
strong for circuits, and the energy scales of the two systems differ. 
In contrast with naturally occuring atoms, artificial atoms can be 
lithographically designed to have specific characteristics, 
such as a large dipole moment or particular transition frequencies. 
That tunability is an important advantage over natural atoms.

Josephson junctions{\bf ---}superconducting grains or electrodes separated by 
an insulating oxide{\bf ---}act like nonlinear inductors in a circuit.
The nonlinearity ensures an unequal spacing between energy levels, so that 
the lowest levels can be addressed using external fields.
Two important energy scales determine the quantum mechanical behavior of a
Josephson-junction circuit: the Josephson coupling energy $E_J$ and the electrostatic
Coulomb energy $E_c$ for a single Cooper pair. 
$E_J=I_c\Phi_0/2\pi$, where $I_c$ denotes the critical current of the junction 
and $\Phi_0=h/2e$ is the magnetic-flux quantum. The charging energy 
$E_c=(2e)^2/2C$ for a Cooper pair, where $C$ is either the capacitance of a Josephson 
junction or an island, depending on the circuit. 
In analogy to the usual position-momentum duality in quantum mechanics,
the phase $\phi$ of the Cooper-pair wave function and the number $n$
of Cooper pairs are conjugate variables and obey the
Heisenberg uncertainty relation $\Delta n \, \Delta \phi \geq 1$.

Box~1 summarizes the four kinds of superconducting qubits
realized in different regimes of $E_J/E_c$.
The charge qubit is in the charge regime $E_c\gg E_J$, where the number $n$ of
Cooper pairs is well defined and the phase $\phi$ fluctuates strongly.
The so-called flux and phase qubits are both in the phase regime $E_c\ll  E_J$,
in which the phase $\phi$ is well defined and $n$ fluctuates strongly. 
And the charge-flux qubit lies in the intermediate regime $E_c\sim  E_J$,
in which charge and phase degrees of freedom play equally important roles.

\vspace{.4cm}\noindent
{\bf\large Charge and charge-flux qubits}

The charge qubit is based on a small superconducting
island known as a Cooper-pair box (CPB), which is coupled to the outside world 
by either one or two weak Josephson junctions and driven by
a voltage source through a gate capacitance (see figue~1a). 
To appreciate how the CPB works, consider a plumbing analogy: 
The box is like a tank that stores water{\bf ---}or in our case, superconducting 
electrons in the form of Cooper pairs.
Those charges can be pushed in and out of the box using a pump (the voltage source) 
that moves the charges through a valve (the Josephson junction) and into the 
superconducting wire that acts as a large reservoir of charges.
Often that one junction is replaced by two that are joined to a segment of 
a superconducting ring and thereby form a symmetric superconducting quantum 
interference device (SQUID). A magnetic flux $\Phi_{\rm ext}$ that pierces 
the SQUID controls the rate at which the Cooper pairs flow into and out of the box.

When the box's offset charge, induced by the gate voltage $V_g$,
is about the same as the charge of a single electron,
only two charge states matter: $|0\rangle$ and $|1\rangle$, which have either zero 
or one extra Cooper pair in the box.
A two-level quantum system thus describes the CPB (see box 2),
and the two energy eigenstates $|\pm\rangle$ are superposition states of
$|0\rangle$ and $|1\rangle$.
The charge qubit can be represented using either the charge states
$\{|0\rangle,|1\rangle\}$ or the eigenstates $\{|+\rangle,|-\rangle\}$.
When the gate-voltage-induced offset charge $n_g$ (in units of $2e$) increases 
from 0, the ground state of the system continuously 
changes from $|0\rangle$ to $|-\rangle$. 
Similarly, the higher energy level $|1\rangle$ becomes $|+\rangle$ for increasing 
$n_g$. At the degeneracy point $n_g=0.5$, where the energy levels for 
$|0\rangle$ and $|1\rangle$ cross, $|\pm\rangle=(|0\rangle \mp |1\rangle)/\sqrt{2}$.

\vspace{.25cm}\noindent \fbox{\parbox{8.4cm}{

{\bf\large Box~2. Cooper-Pair Box}

\vspace{.05cm}\noindent

~~~For the Cooper-pair box (CPB) shown schematically in figure~1a, the Hamiltonian of
the system is
\begin{equation}
H = E_c(n-n_g)^2 - E_J\cos\phi \,,
\end{equation}
where $E_c$ and $E_J$ are the charging and Josephson energies, 
respectively. The phase
drop $\phi$ across the Josephson junction is conjugate to the
number $n$ of extra Cooper pairs in the box. In the
charging regime $E_c\gg E_J$, only the two lowest-lying charge states
of the box, differing by one Cooper pair, are important. 
The gate voltage $V_g$ controls the induced offset charge on the box; 
$n_g=C_gV_g/2e$, where $2e$ is the charge of each Cooper pair and $C_g$ the gate 
capacitance. Around $n_g=1/2$, the system can be described like any two-level
atomic-physics-like system with the reduced Hamiltonian
\begin{equation}
H=\varepsilon(n_g)\;\sigma_z-{1\over 2}E_J\;\sigma_x,
\end{equation}
where $\varepsilon(n_g)=E_c\;(n_g-1/2)$. The Pauli matrices
$\sigma_z=|0\rangle\langle 0|-|1\rangle\langle 1|$ and
$\sigma_x=|0\rangle\langle 1|+|1\rangle\langle 0|$ are defined
in terms of the two basis states corresponding to zero and one extra
Cooper pair in the box.
With a two-junction superconducting quantum interference device, $E_J$ becomes
a very useful, tunable effective coupling:
$E_J(\Phi_{\rm ext})=2E_{J0}\cos(\pi\Phi_{\rm ext}/\Phi_0)$, where $E_{J0}$
is the Josephson coupling energy for each junction, $\Phi_{\rm ext}$ is the 
external magnetic flux, and $\Phi_0$ the flux quantum. }}

\vspace{.2cm}\noindent

\begin{figure*}
\includegraphics[width=4.5in,bbllx=46,bblly=215,bburx=480,bbury=818]
{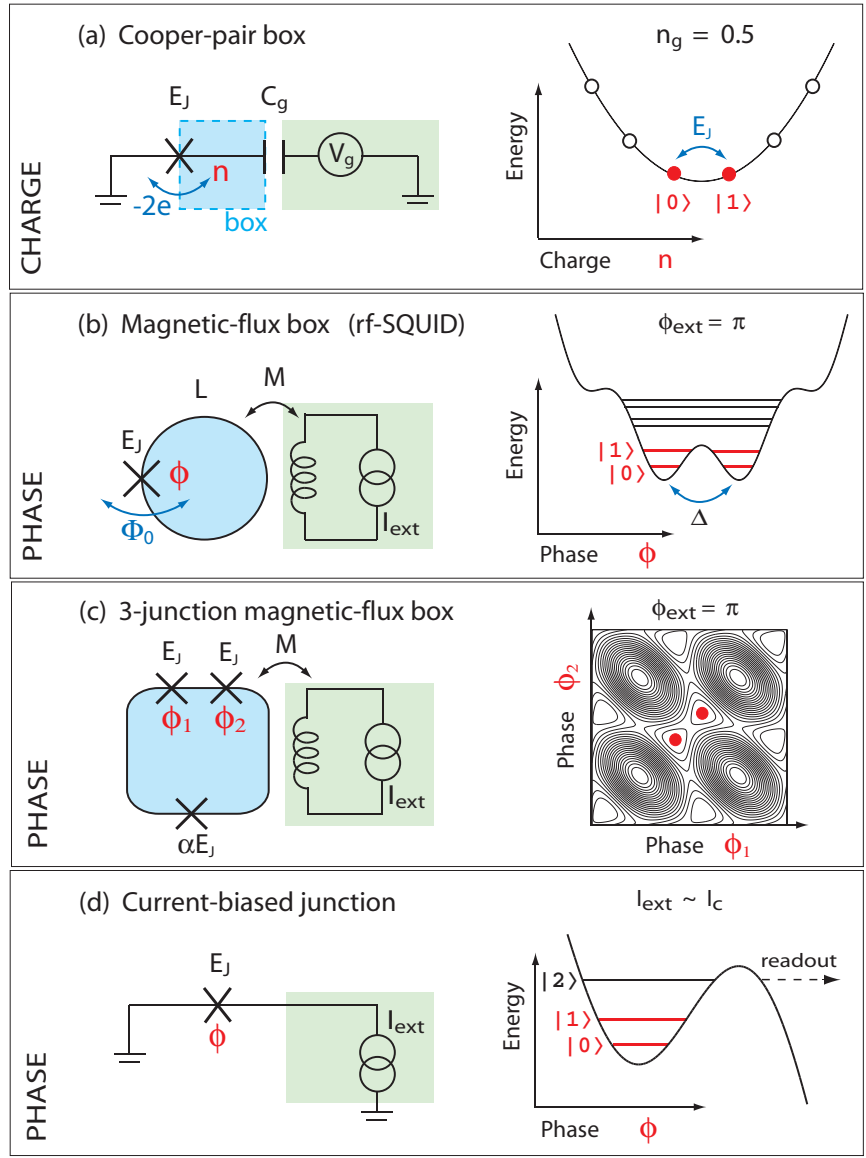}
\caption{{\sc\bf Superconducting qubit circuits} and their potential energy 
diagrams.~{\bf (a)}~A Cooper-pair box (CPB; blue)
is driven by an applied voltage $V_g$ (green) through the gate
capacitance $C_g$ to induce an offset charge $2e\,n_g = C_g V_g$.
A Josephson junction, the barrier denoted by the $\times$, connects the box to 
a wire lead. Each junction has a capacitance and
a Josephson coupling energy $E_J$.
The electrostatic energy of the CPB, $E_c(n-n_g)^2$, where $E_c=(2e)^2/2C$, 
is plotted as a function of the number $n$ of excess electron pairs.
The lowest energy states, $|0\rangle$ and $|1\rangle$ (in red),  are degenerate 
at $n_g=0.5$, and are used as the qubit state basis. Those states are coupled 
via the junction energy $E_J$, which controls the tunneling between them.
{\bf (b)}~A magnetic-flux ``box" (blue) is the magnetic analogue of the 
electrostatic CPB. A magnetic bias simply replaces the electric bias: 
A current-driven magnetic field pierces the box with a strength given by
a mutual inductance $M$. Whereas an electric field prompts stored electron pairs 
to tunnel into or out of the CPB,
a magnetic field pushes magnetic flux quanta
$\Phi_0$ into or out of the superconducting quantum interference device (SQUID) loop.
The adjacent potential energy diagram plots a Josephson energy term
(proportional to $\cos\phi$) and an inductive energy term{\bf ---}proportiona to 
$(\phi-\phi_{\rm ext})^2/2L$, where $L$ is the SQUID's inductance{\bf ---}as a 
function of the phase $\phi$ of the junction.  
The lowest energy states (red) are superpositions of the clockwise and 
counterclockwise supercurrent states $|\!\downarrow\rangle$ and $|\!\uparrow\rangle$
that flow in the SQUID loop;
$\Delta$ here is the tunneling energy between the supercurrent states. 
Those energy states are degenerate when the externally applied magnetic flux 
$\phi_{\rm ext}$ equals $\pi$.
{\bf (c)}~A three-junction flux qubit works like a magnetic flux box, except that 
one of the junctions has a slightly different capacitance and coupling energy.
The contour plot shows the potential energy as a function of two junctions' phases.
The two red dots inside the potential wells correspond to the qubit basis states
$|\!\downarrow\rangle$ and $|\!\uparrow\rangle$.
{\bf (d)}~A current source biases the junction in a phase qubit.
Logic operations can be achieved by driving the qubit with
a microwave field at frequency $(E_1-E_0)/h$. Pulsing the qubit with a microwave 
field at a frequency $(E_2-E_1)/h$ produces a transition
from $|1\rangle$ to $|2\rangle$. One can then read the qubit's state by
measuring the occupation probability of state $|2\rangle$.
}
\end{figure*}

Figure~2 shows the energy spectrum of the CPB for two values of the ratio 
$E_c/E_J$. The regime near $E_c/E_J=5$ is typical of many charge qubits 
studied in the literature. Indeed, quantum coherent
oscillations in circuits were first demonstrated
in this regime~\cite{NEC} and experimental results showed that the qubit can
be well approximated by a two-level system.

When $E_c/E_J = 1$, both charge and flux degrees of freedom play equally
important roles. As shown in figure~2b, the two lowest levels are not well
separated from the higher levels.
Because the qubit now operates in the intermediate regime between charge
and flux, it is often specified as the charge-flux qubit. In contrast with 
the eigenstates of the ideal charge qubit, the charge-flux qubit's two lowest 
eigenstates $|\pm\rangle$ are superpositions of several charge states, 
instead of just $|0\rangle$ and $|1\rangle$.
Thus only $\{|+\rangle,|-\rangle\}$ can be used as the basis states for 
the charge-flux qubit.
This qubit can exhibit coherent oscillations with a long
decoherence time~\cite{Saclay}{\bf ---}on the order of $0.5$~$\mu$s.

\begin{figure}
\includegraphics[width=3.4in,bbllx=88,bblly=235,bburx=545,bbury=481]
{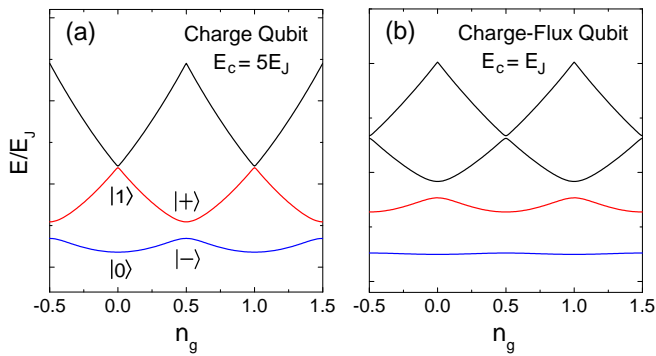}
\caption{{\sc\bf Energy levels of a Cooper-pair box} versus the offset charge
$n_g$ (in units of $2e$, twice the charge of an electron) that is induced by the gate 
voltage. {\bf (a)}~When $E_c/E_J=5$, typical conditions for the charge qubit, the two 
lowest energy levels start to approach each other as the offset charge on the box 
increases from 0 to $0.5$. As $n_g$ slowly increases in that range, eigenstates 
of the two levels change from charge states $|0\rangle$ and $|1\rangle$ to 
$|\pm\rangle$, superpositions of $|0\rangle$ and $|1\rangle$. They become pure 
charge states again at $n_g=1$. When $n_g$ is about $0.5$, the two lowest energy 
levels are well separated from the other levels. 
{\bf (b)}~In the case where $E_c/E_J=1$, the charge and flux degrees of freedom 
play equally important roles.
}
\end{figure}


\vspace{.4cm}\noindent
{\bf\large Flux qubit}

The phase degree of freedom becomes dominant in the so-called flux qubit.
As sketched in figure~1c, the prototypical flux qubit consists of a superconducting 
loop with three junctions, and the Josephson coupling energy is much
larger than the charging energy for each junction.
When a magnetic field is applied through the loop,
a clockwise or counterclockwise supercurrent is induced to decrease
or increase the enclosed flux such that the fluxoid, which combines the Josephson 
phase $\phi$ with the total magnetic flux (both external $\Phi_{\rm ext}$ and 
induced $\Phi_{\rm ind}$), is quantized:
$(\Phi_0/2\pi)\,\phi + \Phi_{\rm ext} + \Phi_{\rm ind} = m\,\Phi_0$\,, 
where $m$ is an integer.
The two circulating supercurrent states can form the basis states for the qubit.
Five years ago, a one-junction flux qubit of the type sketched in figure~1b was 
fabricated and its spectroscopic features were demontrated~\cite{Stony}. 
Flux qubits can come in one- or many-junction flavors;
the one-junction case requires a relatively large loop inductance, 
which makes the qubit more susceptible to magnetic-field noise.

\begin{figure}
\includegraphics[width=3.4in,bbllx=82,bblly=363,bburx=538,bbury=622]
{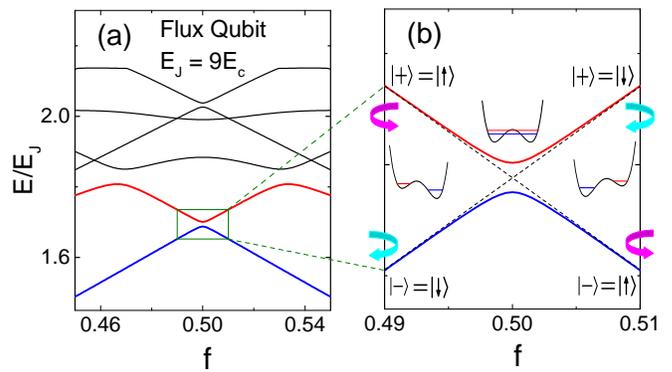}
\caption{
{\sc\bf Energy levels of a three-junction flux qubit} versus reduced magnetic flux
$f=\Phi_{\rm ext}/\Phi_0$. $\Phi_{\rm ext}$ is the external magnetic flux and 
$\Phi_0$ the flux quantum, $h/2e$. 
{\bf (a)}~The energy-level diagram shows only six levels, the two lowest of which 
are used for the qubit.
{\bf (b)}~An enlarged view of those levels near $f=1/2$ (green rectangle) illustrates 
what happens as the reduced magnetic flux varies around that value. 
Away from $f=1/2$, the two eigenstates approach the clockwise- 
and counterclockwise-circulating supercurrent states $|\!\!\downarrow\rangle$ 
and $|\!\!\uparrow\rangle$; at $f=1/2$, they are maximal superpositions of the two 
circulating supercurrent states. The potential energy minimum shifts from the 
right-hand side of the double well for $f<0.5$ to the left-hand side for $f>0.5$, 
where $|\!\!\uparrow\rangle$ becomes the minimum-energy state.
}
\end{figure}

For a description of the energy spectrum that arises in a multi-junction flux-qubit 
system, see figure~3 (figure~1c pictures the corresponding circuit).
In the vicinity of $f=\Phi_{\rm ext}/\Phi_0=0.5$, $|-\rangle$ and $|+\rangle$, 
the two lowest levels (the qubit levels), are well separated from other higher levels, 
and are superpositions of the clockwise and counterclockwise supercurrent states 
$|\!\!\downarrow\rangle$ and $|\!\!\uparrow\rangle$. 
For $f<0.5$, $|-\rangle$ and $|+\rangle$ approach 
$|\!\!\downarrow\rangle$ and $|\!\!\uparrow\rangle$; 
for $f>0.5$, $|-\rangle$ and $|+\rangle$ approach 
$|\!\!\uparrow\rangle$ and $|\!\!\downarrow\rangle$. 
At $f=0.5$, the states are given 
by $|-\rangle=(|\!\!\uparrow\rangle+|\!\!\downarrow\rangle)/\sqrt{2}$ and
$|+\rangle=(|\!\!\uparrow\rangle-|\!\!\downarrow\rangle)/\sqrt{2}$.
As in the case of the charge qubit, one can use either
$\{|\!\!\uparrow\rangle,|\!\!\downarrow\rangle\}$
or $\{|-\rangle,|+\rangle\}$ to equivalently represent the flux qubit.
For the past five years researchers have 
studied the three-junction 
flux qubit, and in 2003 first observed its quantum coherent oscillations~\cite{Delft}.

\vspace{.4cm}\noindent
{\bf\large Phase qubit}

The so-called phase qubit usually uses a large current-biased Josephson 
junction, as pictured in figure~1d. The bias current produces a tilt to the 
Josephson potential; the Josephson potential itself is proportional $\cos\phi$. 
That tilt reduces the number of bound states in the potential-energy well. 
The ratio $E_J/E_c$ is orders of magnitude higher in the phase qubit 
than in other qubit types. 

The circuit's potential energy diagram illustrates a third energy level not 
widely separated from the two lowest levels used for the qubit. The small energy 
spacing means that appreciable qubit-state leakage to that third level can occur. 
However, the problem can at least partly turn into an advantage
when it comes to measuring the phase qubit's actual quantum states.
The state of the third level can easily tunnel out of the potential
well and thus be used for determining the occupation probability of the qubit levels. 
Alternatively, one can read out the qubit state by tilting the potential to allow 
tunneling directly from $|1\rangle$. Independent research groups have 
experimentally demonstrated the quantum coherent oscillations in phase-qubit 
circuits~\cite{KB}.

A phase qubit can also be configured into a circuit similar to what's shown 
for the magnetic-flux box
in figure~1b by biasing the junction using a flux threading the loop instead of 
using a current. Such a flux-biased phase qubit works with levels in a tilted well, 
as depicted in figure~1d.

\vspace{.4cm}\noindent
{\bf\large Coupling qubits}

Two-qubit operations are required for quantum computing. 
A natural way to couple circuit-based qubits to build logic gates is to use 
capacitors and inductors. Figure~4 illustrate a few circuit configurations that 
could do the job. 

Recent experiments have shown quantum coherent oscillations
in two capacitively coupled charge qubits and demonstrated a 
working controlled-NOT (CNOT) gate~\cite{RIKEN}. However, controlling the interbit 
capacitive coupling is a difficult problem~\cite{Averin}.
An alternative is to couple charge qubits via an inductance~\cite{You},
which produces a flux-controllable interbit coupling and can be conveniently used to achieve 
a CNOT gate (see box 3).

\begin{figure}
\includegraphics[width=3.4in,bbllx=16,bblly=326,bburx=567,bbury=770]
{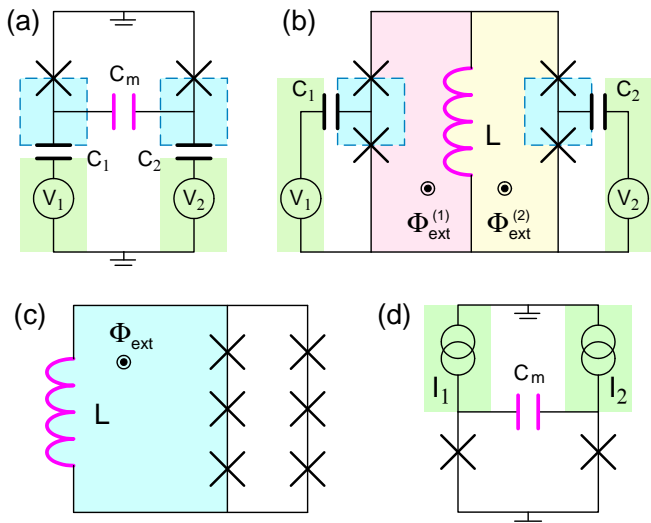}
\caption{{\sc\bf Coupling qubits.}~{\bf (a)}~Two charge qubits (blue)
coupled by a mutual capacitance $C_m$ between the two boxes.
{\bf (b)}~Two charge qubits coupled by a shared inductance $L$; 
each box (blue) is connected to two Josephson junctions instead of one.
{\bf (c)}~Two three-junction flux qubits coupled by the common inductance $L$.
{\bf (d)}~Two phase qubits coupled by the mutual capacitance $C_m$ between them.
}
\end{figure}

An inductance can also couple flux qubits.
Because the Josephson coupling energies in the flux qubits are stronger than
those in the charge qubits, the circulating supercurrents in the qubit loops are larger.
Therefore, a much smaller inductance produces a relatively strong interbit 
coupling~\cite{YNN}. Recently, coupling flux qubits with mutual inductances 
was experimentally realized~\cite{MAJ}.

Two phase qubits can be similarly coupled using a mutual capacitance, an experimental 
achievement made by researchers~\cite{Maryland}. Again,
the lowest two levels for each phase qubit are not widely separated from the third one.
That means that energy levels beyond the qubit are also involved in the coupling
and relatively serious qubit-state leakage can occur for the two-qubit gate.
Achieving controllable interbit coupling is still a challenge in general for any type of qubit.

\vspace{.25cm}\noindent \fbox{\parbox{8.4cm}{

{\bf\large Box~3. Controlled-NOT Gate}

\vspace{.05cm}\noindent

~~~For the two inductively coupled qubits shown in figure~4b,
when the gate voltage is shifted to the degeneracy
point $n_{gi}=1/2$ ($i=1,2$) for each qubit, the Hamiltonian of the system becomes
\[
H=-E^*_{J1}\,\sigma_{x}^{(1)}-E^*_{J2}\,\sigma_{x}^{(2)} +
\chi\,\sigma_{x}^{(1)}\;\sigma_{x}^{(2)}\, ,
\]
where $E^*_{Ji}$ slightly deviates from $E_{Ji}$ and the interbit coupling $\chi$
is controllable via the external fluxes $\Phi_{\rm ext}^{(i)}$.
This $4\times 4$ Hamiltonian has four eigenvalues with corresponding eigenstates
$|+,+\rangle$,$|+,-\rangle$,$|-,+\rangle$, and $|-,-\rangle$,
where $|\pm \rangle=(|0\rangle \mp |1\rangle)/\sqrt{2}$.
The Hamiltonian also has the interesting property that its eigenvalues
change with the interbit coupling, but the corresponding
eigenstates remain unchanged.
Because the energy levels of those four eiegenstates are not equally spaced, 
a microwave field applied to the coupled qubits through either gate capacitance
can be tuned to make transitions only between states $|-,+\rangle$ and $|-,-\rangle$. 
When a $\pi$ pulse from such a field is applied, those states flip to produce a CNOT gate:
$|+,+\rangle\;\longrightarrow\; |+,+\rangle$,
$|+,-\rangle\;\longrightarrow\; |+,-\rangle$,
$|-,+\rangle\;\longrightarrow\; |-,-\rangle$, and
$|-,-\rangle\;\longrightarrow\; |-,+\rangle$. That is, the state
of the second qubit is flipped if the first qubit state is
$|-\rangle$, and the second qubit is not affected if the first
qubit state is $|+\rangle$. (See reference~\cite{You}.) 
}}

\vspace{.2cm}\noindent

\vspace{.4cm}\noindent
{\bf\large Cavity quantum electrodynamics}

A quantized electromagnetic field can coherently exchange
energy with a two-level system, usually in a tiny laser cavity. This energy exchange  
between the field and the system, called Rabi oscillations, occurs at a rate $\nu$ proportional 
to the strength of the system-field coupling.
Among such coherent processes, the most elementary one involves the interaction
of a two-level system with a single photon. The exchange of energy between the system and 
the photon is observable in the ``strong coupling" regime, when the period $1/\nu$
of the Rabi oscillations is much shorter than both the
decoherence time of the two-level system and the average lifetime of the photon
in the cavity. The strong-coupling limit has been achieved for a variety of atoms interacting
with the light field in a cavity and forms the basis
of a subject called cavity quantum electrodynamics (QED).

In principle, any type of two-level quantum system can substitute for the atom, and 
the charge qubit, as a macroscopic quantum system, is a natural candidate. 
Indeed, we and the Yale group proposed schemes to process quantum information by coupling 
a charge qubit with a quantized microwave field. 
One approach took advantage of magnetic coupling through a SQUID loop~\cite{YN}
and the other exploited the control given by the gate voltage and capacitive coupling~\cite{Blais}.

In a more recent experiment, 
Ref.~\onlinecite{Yale} carried out those proposals by using 
a gate capacitor to couple the photon to a Cooper-pair box, an accomplishment coined 
``circuit QED" because it translated the cavity QED concept onto a solid-state chip 
(see Physics Today, November 2004, page 25). 
The researchers reached the strong-coupling regime using a quasi-one-dimensional transmission-line 
resonator. In contrast to cavity QED, where atoms move around and only briefly interact 
with the field, circuit QED uses a charge qubit that is fixed on the chip.
More important, the dipole moment that couples the two-level system with the quantized field 
can be as much as $10^5$ times larger for superconducting charge qubits than for alkali atoms. 

The experiment can take different forms. Other roups have modified it by replacing the cavity 
with a hamonic oscillator formed by a Josephson junction (or SQUID)
and a nanomechanical resonator~\cite{resonator}.
Future ways to exploit superconducting qubits include, for example,
preparing Schr{\"o}dinger cat states of the cavity field by means of its coupling to
a SQUID-based charge qubit, and
the exciting possibility of generating nonclassical photon states
using a superconducting qubit in a microcavity~\cite{Liu}.
Clearly, the experiments are opening new research directions.

\vspace{.4cm}\noindent
{\bf\large Noise and decoherence}

Although superconducting circuits exhibit good quantum coherence,
they still experience significant levels of noise due to 
their coupling to the environment. For charge qubits, the dominant source of decoherence 
is $1/f$ noise, which is presumably due to background charge fluctuations{\bf ---}trapped 
charges in the substrate and oxide layers of the Josephson junctions, for instance.
For flux and phase qubits, $1/f$ noise again seems to be dominant, but its origins 
are less clear. When the CPB operates in the charge-flux regime, the decoherence of the qubit 
is significantly reduced~\cite{Saclay}. 
Moreover, the decoherence can be suppressed at the degeneracy point by tuning 
the magnetic and electric fields~\cite{Saclay} 
so that the influence of both flux and charge noise sources vanishes to first order.

To try to understand the decoherence problem, researchers have used phenomenological theories 
including the spin-boson~\cite{MSS} and spin-fluctuator~\cite{PALA} models
in which a collection of spectrally distributed harmonic oscillators and a set of 
particles that fluctuate randomly in a double-well potential, respectively, describe the noise.
Such models capture some typical features of decoherence in superconducting qubits.
Nevertheless, understanding the microscopic mechanisms of $1/f$ noise requires further 
work{\bf---}for instance, developing microscopic theories beyond phenomenological models.
Such understanding is important not only for quantum computing, but also for revealing
the underlying physics. The problem has proven to be quite difficult, however.

\vspace{.4cm}\noindent
{\bf\large What lies beyond}

Decoherence is a major obstacle to superconducting quantum computing;
the efficient and nondissipative readout of qubit states, however, is also crucial 
and will play a central role in future developments.
Thus, it is still too early to say which type of qubit might win the race of
quantum computing. While unveiling the microscopic mechanism of $1/f$ noise, one could
develop novel methods to actively suppress the effects of the noise.
Also, to increase both decoherence time and readout efficiency of the system,
one can optimize the qubits by varying the circuit parameters and could couple two or three 
qubits with optimal designs; a three-qubit circuit
could be used to test some simple quantum algorithms, such as the Deutsch algorithm, 
one of the simplest that illustrates the nature of quantum parallelism.

So far, all quantum states on the so-called Bloch sphere{\bf ---}a geometrical representation 
of the states of a two-level system{\bf ---} can be addressed; spin-echo 
techniques, borrowed from nuclear magnetic resonance, can reduce the effect of $1/f$ noise; 
and readout efficiency greater than $96\%$ 
and a coherence 
quality factor of approximately $10^5$ can be achieved, albeit not in the same circuit.
When techniques for manipulating two or three qubits become well established,
the next step will be to build circuits with a larger number of qubits,
increased readout efficiency, and lower decoherence.
Such conditions would allow quantum computing with superconducting qubits.

But even if no quantum computing is ever achieved with superconducting circuits,
they still provide researchers with tools to test fundamental quantum mechanics in novel ways.
For example, these artificial atoms can be used to simulate atomic physics using quantum
circuits; researchers have already observed Rabi oscillations and Ramsey interference 
patterns that are manifest during the phase evolution of a superconducting qubit. 
Moreover, the device can also test Bell inequalities, produce Schr{\"o}dinger-cat states, 
and simulate the Einstein-Podolsky-Rosen experiment. The quantum engineering of
macroscopic entangled states will surely play a central role in several future
technologies.

\vspace{.2cm}\noindent

{\it We thank our collaborators and other colleagues, many listed below,
for their valuable contributions.
This work was supported in part by the National Security Agency, Advanced Research and Development 
Activity, Air Force Office of Scientific Research,, NSF, and NSFC (National Natural Science 
Foundation of China).}


\end{document}